\documentclass[paper]{ieice}

\usepackage{cite}
\usepackage{insertfig}
\usepackage{times}
\usepackage{amsmath,amssymb}
\usepackage{url}
\usepackage{enumitem}
\usepackage{amsthm}

\renewcommand{\insertfigsize}{0.4}
\newtheorem*{theorem}{Theorem}
\newtheorem*{remark}{Remark}

\newcommand{\ds}{\displaystyle}

\usepackage{bm}
\newcommand{\vbf}[1]{\boldsymbol{#1}}

\newcommand{\tmp}[1]{{#1}^\mathsf{\!T}}
\newcommand{\inv}[1]{#1^{-1}}
\newcommand{\prob}[1]{\mathbb{P}\left[#1\right]}
\newcommand{\vl}[1]{{\rm Vol}(#1)}

\newcommand{\diff}{{\rm d}}
\newcommand{\diag}[1]{{\rm diag}(#1)}

\newcommand{\ex}[2]{\mathbb{E}_{#1\!}\bigl[#2\bigr]}
\newcommand{\exx}[2]{\mathbb{E}_{#1\!}\Biggl[#2\Biggr]}
\newcommand{\tr}[1]{\mathrm{Tr}\!\left[#1\right]}
\newcommand{\comb}[2]{\binom{#1}{#2}}

\begin{document}

\title{The Wigner's Semicircle Law of Weighted Random Networks}

\authorlist{\authorentry[sakumoto@kwansei.ac.jp]{Yusuke Sakumoto}{m}{kwansei}
   \authorentry[aida@tmu.ac.jp]{Masaki Aida}{f}{tmu-univ}
}

\affiliate[kwansei]{Kwansei Gakuin University, 2-1 Gakuen, Sanda, Hyogo 669-1337, Japan}
\affiliate[tmu-univ]{Tokyo Metropolitan University, 6-6 Asahigaoka, Hino, Tokyo 191-0065, Japan}

\maketitle

\begin{summary}
The spectral graph theory provides an algebraical approach to
investigate the characteristics of weighted networks using the
eigenvalues and eigenvectors of a matrix~(e.g., normalized Laplacian
matrix) that represents the structure of the network.  However, it is
difficult for large-scale and complex networks~(e.g., social network)
to represent their structure as a matrix correctly.  If there is a
universality that the eigenvalues are independent of the detailed
structure in large-scale and complex network, we can avoid the
difficulty.  In this paper, we clarify the Wigner's Semicircle Law for
weighted networks as such a universality.  The law indicates that the
eigenvalues of the normalized Laplacian matrix for weighted networks
can be calculated from the a few network statistics~(the average
degree, the average link weight, and the square average link weight)
when the weighted networks satisfy the sufficient condition of the
node degrees and the link weights.
\end{summary}


\section{Introduction}

Many networks~(e.g., railway network, social network, the Internet,
and airport network) are often modeled as weighted networks that are
composed of nodes and weighted
links~\cite{Shi16:Social,Feng17:Complex,Sun17:Airport}.  The weighting
of links is important to model a network, but it is hard for
large-scale and complex networks such as social network.  In the
social network, nodes and links correspond to persons and their
acquaintance relationships, respectively.  In order to correctly give
the weight for each link in the social network, the strength of the
relationships among people should be accurately estimated from a huge
amount of personal data~(e.g., communication histories in mobile phone
and social media).  Due to the problem of the privacy and the
computational complexity, it is unrealistic to gather such personal
data, and calculate the strength of the relationships accurately.

The spectral graph theory provides an algebraical approach to
investigate the characteristics of weighted networks using the
eigenvalues and the eigenvectors of a matrix~(e.g., normalized
Laplacian matrix) that represents the structure of the
network~\cite{Chung97:Spectral,Spielman07:Spectral}.  In particular,
the eigenvalues are important to understand the characteristics
related to the entire network on the basis of the spectral graph
theory. In~{\cite{Sakumoto19:IEICE}}, we have been clarified that the
eigenvalue distribution of the normalized Laplacian matrix affects the
information dissemination speed throughout the social network.  In
general, the eigenvalues of the matrix are calculated from all the
matrix elements.  For the matrix representing the structure of a
weighted network, its elements are determined by the links weights.
Hence, the weighting of the links is required to apply the spectral
graph theory for weighted networks.  This would be the barrier to
apply the spectral graph theory for large-scale and complex networks
due to the above-mentioned problems.  However, the weighting of the
links is avoidable if there is a universality that the eigenvalues 
are independent of the detailed structure~(e.g., the weight of each
link) of large-scale and complex networks.  Therefore, the finding of
such a universality of the eigenvalues expands the applicable region of
the spectral graph theory.

The random matrix theory discusses a universality of the eigenvalues
if the elements of the matrix are given by random
variables~\cite{Wigner58:Law,Vasiliki02:RMT,Chung03:Spectra}.
In~\cite{Chung03:Spectra}, Chung et al. have been analyzed the random
matrix corresponding to the normalized Laplacian matrices for
unweighted networks, and have clarified the universality~(the Wigner's
semicircle law) that the eigenvalues of the normalized Laplacian
matrix follow the semicircle distribution.  To our knowledge, no study
has clarified a universality of the eigenvalues for weighted networks.

In this paper, we clarify the universality~(the Wigner's semicircle
law) of the eigenvalues for weighted networks on the basis of the
discussion of ~\cite{Chung03:Spectra}.  The clarified universality
indicates that the eigenvalues of the normalized Laplacian matrix for
weighted networks can be calculated from the a few network
statistics~(the average degree, the average link weight, and the
square average link weight) when the weighted networks satisfy the
sufficient condition of the node degrees and the link weights.  Using
some numerical examples, we confirm the validity of the sufficient
condition.

This paper is organized as follows.  In Sect.~\ref{sec:preliminary},
we describe the spectral graph theory and the random matrix theory for
weighted networks.  In Sect.~\ref{sec:analysis}, we prove the Wigner's
semicircle law for weighted networks on the basis of the random matrix
theory.  Section~\ref{sec:example} shows some numerical examples.
Finally, in Sect.~\ref{sec:conclusion}, we conclude this paper and
discuss the future work.

\section{Preliminary}
\label{sec:preliminary}
\subsection{Spectral Graph Theory}

In the spectral graph theory, the structure of a network is
represented by the matrix, and its characteristics is investigated
using the eigenvalues and the eigenvectors of the matrix.  In this
section, we describe the spectral graph theory for weighted networks.

We denote a weighted network by $G = (V, E, \vbf{w})$ where $V$ and
$E$ are the sets of nodes and links, respectively. Let $n$ be the
number of nodes in $G$.  The link between nodes $i$ and $j$ is denoted by
$(i, j)$.  Link $(i, j)$ has the weight $w(i, j)$ where $w(i, j) =
w(j, i)$ and $w(i, j) > 0$.  Let $\partial i$ be the set of adjacency
nodes of node $i$.  The weighted degree $d_i$ of nodes $i$ is defined by
\begin{align}
  d_i := \sum_{j \in \partial i} w(i, j).
\end{align}

To represent the structure of links and nodes in $G$, there are
adjacency matrix $\vbf{A}$ and degree matrix $\vbf{D}$, respectively.
The $(i, j)$-th element $A(i, j)$ of adjacency matrix $\vbf{A}$ is
defined by
\begin{align}
   A(i, j) := 
   \begin{cases}
     w(i, j)   &  \mathrm{if} \, (i,j) \in E \\
     0         &  \mathrm{otherwise}
   \end{cases}  
   \label{eq:def_A}.
\end{align}
Degree matrix $\vbf{D}$ is defined by
\begin{align}
  \vbf{D} := \diag{d_i}_{1 \le i \le n}.
\end{align}
To represent the both structure of nodes and links in $G$, normalized
Laplacian matrix $\vbf{N}$ is often used. Normalized Laplacian matrix
$\vbf{N}$ is defined by
\begin{align}
  \vbf{N} &:= \vbf{I} - \vbf{D}^{-1/2}\vbf{A}\vbf{D}^{-1/2} \label{eq:def_N},
\end{align}
where $\vbf{I}$ is the identity matrix.

Since normalized Laplacian matrix $\vbf{N}$ is symmetric~$(\vbf{N} =
\tmp{\vbf{N}})$, its eigenvalue $\lambda_l$~($l = 1, ..., n$) are real
numbers.  Let $\vbf{q}_l$ be the eigenvector of eigenvalue $\lambda_l$
where $\tmp{\vbf{q}_l}\vbf{q}_l = 1$.  We assign a number to
$\lambda_l$ in ascending order, and hence $\lambda_l$ means the $l$-th
minimum eigenvalue of $\vbf{N}$.  The range of the eigenvalues is given by 
\begin{align}
  0 = \lambda_1 < \lambda_2 \le ... \le \lambda_n < 2.
  \label{eq:relation_lambda}
\end{align}
If $\lambda_2 > 0$, weighted network $G$ is connected~(i.e., there is
at least one path between every pair of nodes).  Then, weighted
network $G$ is not bipartite graph if $\lambda_2 < 2$.  We define
spectral radius $r$ by $r := \max_{2 \le l \le n} |1-\lambda_l|$.
Spectral radius $r$ is in $0 < r < 1$ because $0< \lambda_l < 2$ for
$2 \le l \le n$.  Eigenvector $\vbf{q}_1$ of minimum eigenvalue
$\lambda_1$ is given by
\begin{align}
  \vbf{q}_1 = \frac{1}{\sqrt{\vl{G}}} \tmp{(\sqrt{d_1}, \sqrt{d_2}, ... , \sqrt{d_n})},
\end{align}
where $\vl{G}$ is defined by
\begin{align}
   \vl{G} := \sum_{i \in G} d_i.
\end{align}
Since eigenvector $\vbf{q}_l$ is the orthonormal basis, matrix
$\vbf{Q} = (\vbf{q}_l)_{1 \le k \le n}$ is the orthogonal
matrix~($\inv{\vbf{Q}} = \tmp{\vbf{Q}}$).

Using $\vbf{Q}$ and $\vbf{\Lambda} = \diag{\lambda_l}_{1 \le k \le n}$,
normalized Laplacian matrix $\vbf{N}$ is given by
\begin{align}
  \vbf{N}  = \vbf{Q}\vbf{\Lambda}\tmp{\vbf{Q}} = \sum_{l=1}^n \lambda_l\,\vbf{q}_l\,\tmp{\vbf{q}}_l.
  \label{eq:spectral_form_N}
\end{align}
From the above equation, $\vbf{N}$ is determined by eigenvalues
$(\lambda)_{1 \le l \le n}$ and eigenvectors $(\vbf{q}_l)_{1 \le l \le
  n}$.  Hence, weighted network $G$ can be analyzed not only with
$\vbf{N}$ but also with their eigenvalues and eigenvectors.  The
spectral graph theory provides an algebraical analysis method of $G$
with eigenvalues $\lambda_l$ and eigenvectors $\vbf{q}_l$ of
$\vbf{N}$.  In particular, eigenvalues $\lambda_l$ is important to
understand the statistical characteristics related to the whole of
weighted network $G$.  In order to calculate eigenvalues $\lambda_l$,
in general, all elements $N(i, j)$ must be given correctly.  However,
if there is a useful universality of eigenvalues $\lambda_l$, we can
investigate the statistical characteristics of weighted network $G$
without all elements $N(i, j)$.

\subsection{Random Matrix Theory}

The random matrix theory focuses on random matrices that the elements
are given by random variables, and clarifies that a universality of
the eigenvalues appears if the matrix size approaches to infinity.
When the links and their weights in weighted network $G$ are randomly
given with a stochastic rule, elements $N(i, j)$ of normalized
Laplacian matrix $\vbf{N}$ become random variables, and $\vbf{N}$ can
be treated as a random matrix.  Note that the universality of
eigenvalues $\lambda_l$ of $\vbf{N}$ corresponds to a characteristic of
the statistical ensemble of the random networks generated with the
same stochastic rule.  Hence, clarifying such a universality
contributes the growth of the statistical mechanics on networks.

In~\cite{Chung03:Spectra}, the link between nodes $i$ and $j$ in
unweighted networks is randomly generated using the stochastic rule
with random variable $L_{ij}$.  If $L_{ij} = 0$, there is no link
between nodes $i$ and $j$.  On the other hand, if $L_{ij} = 1$, the link
exists between nodes $i$ and $j$.  We denote probability
$\prob{L_{ij} = 1}$ by $p_{ij}$, which is given by
\begin{align}
  p_{ij} = \rho\,\omega_i\,\omega_j,
  \label{eq:prob_link}
\end{align}
where $\omega_i > 0$ and $\rho = 1/\sum_{i \in V} \omega_i$.  Using
Eq.~\eqref{eq:prob_link}, the expectation of the node $i$'s degree is
given by $\omega_i$.  Let $\omega_{\rm avg}$, $\omega_{\rm min}$, and
$\omega_{\rm max}$ be the average degree, the minimum degree, and the
maximum degree.  These are defined by $\omega_{\rm avg} := 1/n \sum_{i
  \in V} \omega_i$, $\omega_{\rm min} := \inf_{i \in V} \omega_i$, and
$\omega_{\rm max} := \sup_{i \in V} \omega_i$, respectively.  We can
write $\rho$ by $\rho = 1/(n\,\omega_{\rm avg})$.  Similarly
to~\cite{Chung03:Spectra}, we assume that $\omega_{\rm max}^2 <
1/\rho$ so that $p_{ij} \le 1$.  As the scale of networks becomes
large, the degrees of nodes are likely to become large.  Hence,
similarly to~\cite{Chung03:Spectra}, we assume that $\omega_{\rm min}$
diverges to infinity as $n \rightarrow \infty$.  Using the above
stochastic rule, special networks~(e.g., networks with no links) are
rarely generated.  However, the probability generating such special
networks is very small, and hence this is no problem to discuss a
universality appearing when $n \rightarrow \infty$.

In weighted network $G$, the weight of link $(i, j)$ is randomly set
using the stochastic rule with random variable $W$.  In the stochastic
rule, random variable $W$ follows the conditional probability density
function $f_{W \mid L_{ij}}(w \mid l)$, which is defined by
\begin{align}
  f_{W \mid L_{ij}}(w \mid l) \,\diff w := \prob{ w \le W \le w + \diff w \mid L_{ij} = l}.
\end{align}
If $L_{ij} = 0$~(i.e., link $(i, j)$ does not exist), $W$ is always
$0$, and hence $f_{W \mid L_{ij}}(w \mid 0) = \delta(w)$ where
$\delta(x)$ is the Dirac delta function.  On the contrary, if $L_{ij}
= 1$, $W > 0$.  Since link weights in actual networks cannot be
infinity, we assume that $W$ is bounded.  Using a finite value $w_{\rm
  max}$, $W \le w_{\rm max}$.  For the sake of convenience, we write
\begin{align}
  p_W(w) := f_{W \mid L_{ij}}(w \mid 1).
\end{align}
Let $\ex{p_W}{W^m}$ be the $m$-th moment of $W$ with the condition
$L_{ij} = 1$. $\ex{p_W}{W^m}$ is defined by
\begin{align}
  \ex{p_W}{W^m} &:= \int_0^{w_{\rm max}} w^m \,p_W(w) \,\diff w.
\end{align}
Even if the weights of all links in $G$ are divided by $w_{\rm max}$,
normalized Laplacian matrix $\vbf{N}$ is invariant.  Hence, without
loss of generality, we assume that $w_{\rm max} = 1$.  With this
assumption, $\ex{p_W}{W^m}$ has the following properties:
(a)~$\ex{p_W}{W^l} \le \ex{f_W}{W^m}$ for $l > m$, and (b)
$\ex{p_W}{W^m} \le 1$ since $\ex{p_W}{W^0} = 1$.

Following the above stochastic rules, not only elements $N(i, j)$ but
also eigenvalues $\lambda_l$~($l = 2, ..., n$) of the normalized
Laplacian matrix $\vbf{N}$ for the weighted network $G$ become random
variables depending on the set of random variables~$\vbf{\Gamma} =
(\vbf{L}, W)$ where $\vbf{L} = (L_{ij})_{(i,j) \in V^2}$.
Let $\Lambda$ be the random variable for eigenvalue $\lambda$ of
$\vbf{N}$.
We denote the conditional eigenvalue density of eigenvalues
$\lambda_l$~($l = 2, ..., n$) of $\vbf{N}$ by $f_{\vbf{\Lambda} \mid
  \vbf{\Gamma}}^{(n)}(\lambda \mid \gamma)$, which is defined by
\begin{align}
  f_{\Lambda \mid \vbf{\Gamma}}^{(n)}(\lambda \mid \vbf{\gamma}) := \frac{1}{n-1} \sum_{l = 2}^n \delta\left( \lambda - \lambda_l \right),
\end{align}
where $\lambda_l$ is the function of $\vbf{\gamma}$.
Since eigenvalues $\lambda_l$ vary stochastically, conditional
eigenvalue density $f_{\Lambda \mid \vbf{\Gamma}}^{(n)}(\lambda \mid
\vbf{\gamma})$ is also a random variable.
Note that $f_{\Lambda \mid \vbf{\Gamma}}^{(n)}(\lambda \mid
\vbf{\gamma})$ can be treated as a conditional probability density
function because $\int_{\lambda_2}^{\lambda_n} f_{\Lambda \mid
  \vbf{\Gamma}}^{(n)}(\lambda \mid \vbf{\gamma}) \,\diff \lambda = 1$.
Using probability $\prob{\vbf{\Gamma} = \vbf{\gamma}}$, eigenvalue
density $f_{\Lambda}^{(n)}(\lambda)$ of $\vbf{N}$ is given by
\begin{align}
  f_{\Lambda}^{(n)}(\lambda) &= \ex{\prob{\vbf{\Gamma}}}{f_{\Lambda \mid \vbf{\Gamma}}^{(n)}(\lambda \mid \vbf{\gamma})} \nonumber\\
  &= \frac{1}{n-1} \sum_{\vbf{\gamma}} \prob{\vbf{\Gamma} = \vbf{\gamma}} \sum_{l = 2}^n \delta\left( \lambda - \lambda_l \right).
\end{align}
Since $\int_{\lambda_2}^{\lambda_n} f_{\Lambda}^{(n)}(\lambda) \,\diff
\lambda = 1$, $f_{\Lambda}^{(n)}(\lambda)$ can be also treated as a
probability density function.
Then, we denote the $m$-th moment for $1-\Lambda$ using
$f_{\Lambda}^{(n)}(\lambda)$ by
$\ex{f^{(n)}_{\Lambda}}{(1-\Lambda)^m}$, which is defined by
\begin{align}
  \ex{f^{(n)}_{\Lambda}}{(1-\Lambda)^m} &:= \int_{\lambda_2}^{\lambda_n} (1-\lambda)^m f_{\Lambda}^{(n)}(\lambda) \, \diff \lambda \nonumber\\
                                        &= \frac{1}{n-1} \exx{\prob{\vbf{\Gamma}}}{\sum_{l = 2}^n (1-\lambda_l)^m }.
  \label{eq:eigen_moment}
\end{align}

Using the approach of the random matrix theory, previous
work~\cite{Chung03:Spectra} has clarified the universality (the
Wigner's semicircle law) that eigenvalue density
$f_{\Lambda}^{(n)}(\lambda)$ for unweighted networks follows a certain
distribution~(i.e., semicircle distribution).
If such a universality exists even for weighted network $G$,
eigenvalue density $f_{\Lambda}^{(n)}(\lambda)$ can be obtained
without giving link weights $w(i, j)$ correctly.  This allows the
analysis of $G$ based on spectral graph theory even if $G$ is a
large-scale and complex network such as social network.

\section{Wigner's Semicircle Law of Weighted Network $G$}
\label{sec:analysis}

In this section, we probe the Wigner's semicircle law for weighted
network $G$ on the basis of the discussion in~\cite{Chung03:Spectra}.
The Wigner's semicircle law for $G$ is as follows:
\begin{theorem}
  If weighted network $G$ satisfies degree condition
  \begin{align}
    \omega_{\rm min}^2 \gg \frac{\omega_{\rm avg}}{\ex{p_W}{W^2}},
    \label{eq:cond_degree}
  \end{align}
  $f_{\Lambda}^{(n)}(\lambda)$ of normalized Laplacian matrix
  $\vbf{N}$ converges to semicircle distribution
  $\tilde{f}_{\Lambda}(\lambda)$ as $n \rightarrow \infty$.
  Semicircle distribution $\tilde{f}_{\Lambda}(\lambda)$ is given by
\begin{align}
  \tilde{f}_{\Lambda}(\lambda) = \left\{
  \begin{array}{cl}
    \ds \frac{2}{\pi\,\tilde{r}^{2}} \sqrt{\tilde{r}^2-(1-\lambda)^2} & 1-\tilde{r} < \lambda < 1+\tilde{r} \\  
    0 & \mathrm{otherwise}
    \label{eq:semicircle}
  \end{array}
  \right.,
\end{align}
where $\tilde{r}$ is the limit value of spectral radius $r$ as $n \rightarrow \infty$, and is given by
\begin{equation}
  \tilde{r} = \frac{2}{\sqrt{\omega_{\rm avg}}}\frac{\sqrt{\ex{p_W}{W^2}}}{\ex{p_W}{W}}.
  \label{eq:radius}
\end{equation}
\end{theorem}

\begin{remark}
  When the links in $G$ is not weighted, $W$ is always $1$ if $L_{ij}
  = 1$, and hence $\ex{p_W}{W} = \ex{p_W}{W^2} = 1$.  By substituting
  them into Eqs.~\eqref{eq:cond_degree} and \eqref{eq:radius}, we
  obtain the Wigner's semicircle law for unweighted networks shown
  in~\cite{Chung03:Spectra}.  Therefore, it can be generalized to
  weighted network $G$.
\end{remark}

\begin{proof}

If eigenvalue density $f_{\Lambda}^{(n)}(\lambda)$ is given by
semicircle distribution $\tilde{f}_{\Lambda}(\lambda)$, 
even moment $\ex{\tilde{f}_{\Lambda}}{(1-\Lambda)^{2m}}$ and 
odd moment $\ex{\tilde{f}_{\Lambda}}{(1-\Lambda)^{2m+1}}$ for
$1-\Lambda$ are given by
\begin{align}
  \ex{\tilde{f}_{\Lambda}}{(1-\Lambda)^{2m}}   &= \int_{1-\tilde{r}}^{1+\tilde{r}} (1-\lambda)^{2m} \tilde{f}_{\Lambda}(\lambda) \,\diff \lambda \nonumber\\
                          &= \left(\frac{\tilde{r}}{2}\right) ^{2m} \frac{(2\,m) !}{m !\,(m+1)!} \label{eq:wigner_even_moment}, \\
  \ex{\tilde{f}_{\Lambda}}{(1-\Lambda)^{2m+1}} &= \int_{1-\tilde{r}}^{1+\tilde{r}} (1-\lambda)^{2m+1} \tilde{f}_{\Lambda}(\lambda) \,\diff \lambda \nonumber\\
                          &= 0 \label{eq:wigner_odd_moment}.
\end{align}
Since $\ex{\tilde{f}_{\Lambda}}{\Lambda} = 1$, the above moments for
$1-\Lambda$ correspond to the central moments of $\Lambda$.
The followings are equivalent:
\begin{enumerate}
  \item As $n \rightarrow \infty$, eigenvalue density
  $f_{\Lambda}^{(n)}(\lambda)$ converges to semicircle distribution
    $\tilde{f}_{\Lambda}(\lambda)$.
  \item As $n \rightarrow \infty$, even moment
    $\ex{\tilde{f}_{\Lambda}}{(1-\Lambda)^{2m}}$ and odd moment
    $\ex{\tilde{f}_{\Lambda}}{(1-\Lambda)^{2m+1}}$ converge to
    Eqs.~\eqref{eq:wigner_even_moment} and
    \eqref{eq:wigner_odd_moment}, respectively.  
\end{enumerate}
In order to prove the Wigner's semicircle law for weighted network
$G$, we show that 2. is fulfilled if the degree
condition~\eqref{eq:cond_degree} is satisfied.

For the sake of convenience, we use matrix $\vbf{M}$ removing the
effect of minimum eigenvalue $\lambda_1$ from normalized Laplacian
matrix $\vbf{N}$.  Matrix $\vbf{M}$ is defined by
\begin{align}
  \vbf{M} &:= \sum_{l=2}^n (1-\lambda_l)\,\vbf{q}_l\,\tmp{\vbf{q}}_l = \vbf{I} - \vbf{N} - \vbf{q}_1\,\tmp{\vbf{q}}_1\nonumber\\
  &= \vbf{D}^{-1/2}\vbf{A}\vbf{D}^{-1/2} - \frac{1}{\vl{G}}\vbf{D}^{1/2}\vbf{K}\vbf{D}^{1/2},
  \label{eq:M}
\end{align}
where $\vbf{K}$ is the matrix where all elements are given by $1$, and
corresponds to the adjacency matrix for the complete graph including
the self-loops at all nodes.
Matrix $\vbf{M}$ has $n-1$ nonzero eigenvalues, and $l$-th largest
eigenvalue is given by $1-\lambda_l$.  Hence
\begin{align}
  \tr{\vbf{M}^m} = \sum_{l = 2}^{n} (1 - \lambda_l)^m.
\end{align}
By substituting the above equation into Eq.~\eqref{eq:eigen_moment}, we obtain
\begin{align}
  \ex{f^{(n)}_{\Lambda}}{(1-\Lambda)^m} = \frac{1}{n-1} \ex{\prob{\vbf{\Gamma}}}{\tr{\vbf{M}^m}}.
  \label{eq:trace}
\end{align}

Since $\ex{p_W}{W}$ is finite, weighted degree $d_i$ of the node $i$
converges to its expected value $\ex{\prob{\vbf{\Gamma}}}{d_i}$ as $n
\rightarrow \infty$.  $\ex{\prob{\vbf{\Gamma}}}{d_i}$ is given by
\begin{align}
  \ex{\prob{\vbf{\Gamma}}}{d_i} &= \sum_{j \in V} \left( p_{ij}\ex{f_{W | L_{ij}}}{W \mid L_{ij} = 1} \right. \nonumber\\
  & \hspace{1.5cm} \left. + (1-p_{ij}) \ex{f_{W | L_{ij}}}{W \mid L_{ij} = 0} \right) \nonumber\\
  &= \sum_{j \in V}  p_{ij}\ex{p_W}{W} \nonumber\\
  &= \ex{p_W}{W} \sum_{j \in V}  \rho \,\omega_i \, \omega_j \nonumber\\
  &= \ex{p_W}{W} \omega_i.
  \label{eq:ext_weighted_degree}
\end{align}
Note that we used $p_{ij} = \rho\,\omega_i\omega_j$ to derive the above equation.
By substituting $\ex{\prob{\vbf{\Gamma}}}{d_i}$ into Eq.~\eqref{eq:M},
we obtain matrix $\vbf{C}$, which is given by
\begin{align}
  \vbf{C} = \frac{1}{\ex{p_W}{W}}\vbf{\Omega}^{-1/2}\vbf{A}\vbf{\Omega}^{-1/2} - \rho\,\vbf{\Omega}^{1/2}\vbf{K}\vbf{\Omega}^{1/2},
\end{align}
where $\vbf{\Omega} = \diag{\omega_i}_{1 \le i \le n}$.  As $n
\rightarrow \infty$, matrix $\vbf{M}$ converges to matrix $\vbf{C}$
since $d_i = \ex{\prob{\vbf{\Gamma}}}{d_i}$.
Since the Wigner's semicircle law discusses the limit theorem where $n \rightarrow \infty$, 
there is no problem if we prove it using $\vbf{C}$ instead of $\vbf{M}$.

Element $C(i, j)$ of matrix $\vbf{C}$ is a random variable, and is given by
\begin{align}
   C(i, j) = 
   \begin{cases}
    \ds \frac{w_{ij}}{\ex{p_W}{W}\sqrt{\omega_i\omega_j}} - \rho\sqrt{\omega_i\omega_j} &  \mathrm{if} \, L_{ij} = 1\\
     - \rho\sqrt{\omega_i\omega_j}          &  \mathrm{otherwise}
   \end{cases}  
   \label{eq:def_C}.
\end{align}
Let $\ex{\prob{\vbf{\Gamma}}}{C^m(i,j)}$ be the $m$-th moment of $C(i,
j)$.  Specifically, 1st moment $\ex{\prob{\vbf{\Gamma}}}{C(i,j)}$ is
given by
\begin{align}
  &\ex{\prob{\vbf{\Gamma}}}{C(i, j)} \nonumber\\
  & \hspace{0.3cm} = p_{ij} \exx{f_{W | L_{ij}}}{\frac{W}{\ex{p_W}{W} \sqrt{\omega_i\omega_j}} - \rho\sqrt{\omega_i\omega_j} \,\Biggl|\Biggr.\, L_{ij} = 1} \nonumber\\
  & \hspace{2.5cm} + (1-p_{ij}) \exx{f_{W | L_{ij}}}{\rho\sqrt{\omega_i\omega_j} \,\Biggl|\Biggr.\, L_{ij} = 0}  \nonumber\\
  & \hspace{0.3cm} = 0.
\end{align}
For $m \ge 2$, $m$-th moment $\ex{\prob{\vbf{\Gamma}}}{C(i, j)^m}$ is
bounded by
\begin{align}
   &\ex{\prob{\vbf{\Gamma}}}{C^m(i, j)} \nonumber\\
  & = p_{ij} \exx{f_{W | L_{ij}}}{\!\!\left(\frac{W}{\ex{p_W}{W} \sqrt{\omega_i\omega_j}} - \rho\sqrt{\omega_i\omega_j}\right)^m \!\Biggl|\Biggr. L_{ij} = 1} \nonumber\\
  & \hspace{2.1cm} + (1-p_{ij}) \exx{f_{W | L_{ij}}}{\left(\rho\sqrt{\omega_i\omega_j}\right)^m \!\Biggl|\Biggr. L_{ij} = 0}  \nonumber\\
  &= p_{ij}\sum_{l = 0}^{m-2} \comb{m}{l} \frac{(-p_{ij})^l\ex{p_W}{W^{m-l}}}{\ex{p_W}{W}^{m-l} (\omega_i\omega_j)^{m/2}} \nonumber\\
  &= \rho \sum_{l = 0}^{m-2} \comb{m}{l} \frac{(-\rho)^l\ex{p_W}{W^{m-l}}}{\ex{p_W}{W}^{m-l} (\omega_i\omega_j)^{m/2-l-1}}   \label{eq:moment_cij_pre}\\
  &= \frac{\rho\,\ex{p_W}{W^m}}{\ex{p_W}{W}^m (\omega_i\omega_j)^{m/2-1}} \left(1+o(1)\right) \nonumber\\
  &\le \frac{\rho\,\ex{p_W}{W^m}}{\ex{p_W}{W}^m\omega_{\rm min}^{m-2}} \left(1+o(1)\right)
  \label{eq:moment_cij}.
\end{align}
To derive the above equation, we left the term with $l = 0$ in the sum
since $\rho = 1/(n\,\omega_{\rm avg}) = 1/\Omega(n) = o(1)$.
Note that $\Omega(f(n))$ is a function that increases faster than or
equal to $f(n)$ as $n \rightarrow \infty$.  Then, we use $f(n) = o(1)$ as 
\begin{align}
  \lim_{n \rightarrow \infty} f(n) = 0.
\end{align}

$m$-th moment $\tr{\vbf{C}^{m}}$ is given by
\begin{align}
  \tr{\vbf{C}^m} &= \sum_{\vbf{v}_m \in \Phi_{n, m}} C(v_1, v_2) C(v_2, v_3) \,...\, C(v_m, v_1)\nonumber\\
  &= \sum_{\vbf{v}_m \in \Phi_{n, m}} \prod_{l = 1}^{h(\vbf{v}_m)} C(e_l)^{m_l},
  \label{eq:tr_cm}
\end{align}
where $\vbf{v}_m = (v_1, v_2, v_3, ..., v_m)$ represents a cycle with
length $m$ in the complete graph with $n$ nodes, and $\Phi_{n, m}$ is
the set of the cycles with length $m$.  In cycle $\vbf{v}_m$,
$h(\vbf{v}_m)$ is the number of disjoint links, $e_l$ is the $l$-th
link, and $m_l$ is the occurrence number of link $e_l$.  In
particular, $m_l$ is satisfied with
\begin{align}
  \sum_{l = 1}^{h(\vbf{v}_m)} m_l = m.
\end{align}
where $(i, j)$ and $(j, i)$ are treated as the same link in a cycle
when counting $m_l$ since $\vbf{C}$ is a symmetric matrix.

Figure~\ref{fig:complete_graph} shows the example of the complete graph with $4$ nodes.
For example, $\Phi_{4, 6}$ includes $\vbf{v}_6 = (1, 2, 3, 1, 2, 3)$.
In cycle $\vbf{v}_6$, $h(\vbf{v}_6) = 3$, $e_2 = (2, 3)$ and $m_2 = 2$.

\insertPDFfig[.15]{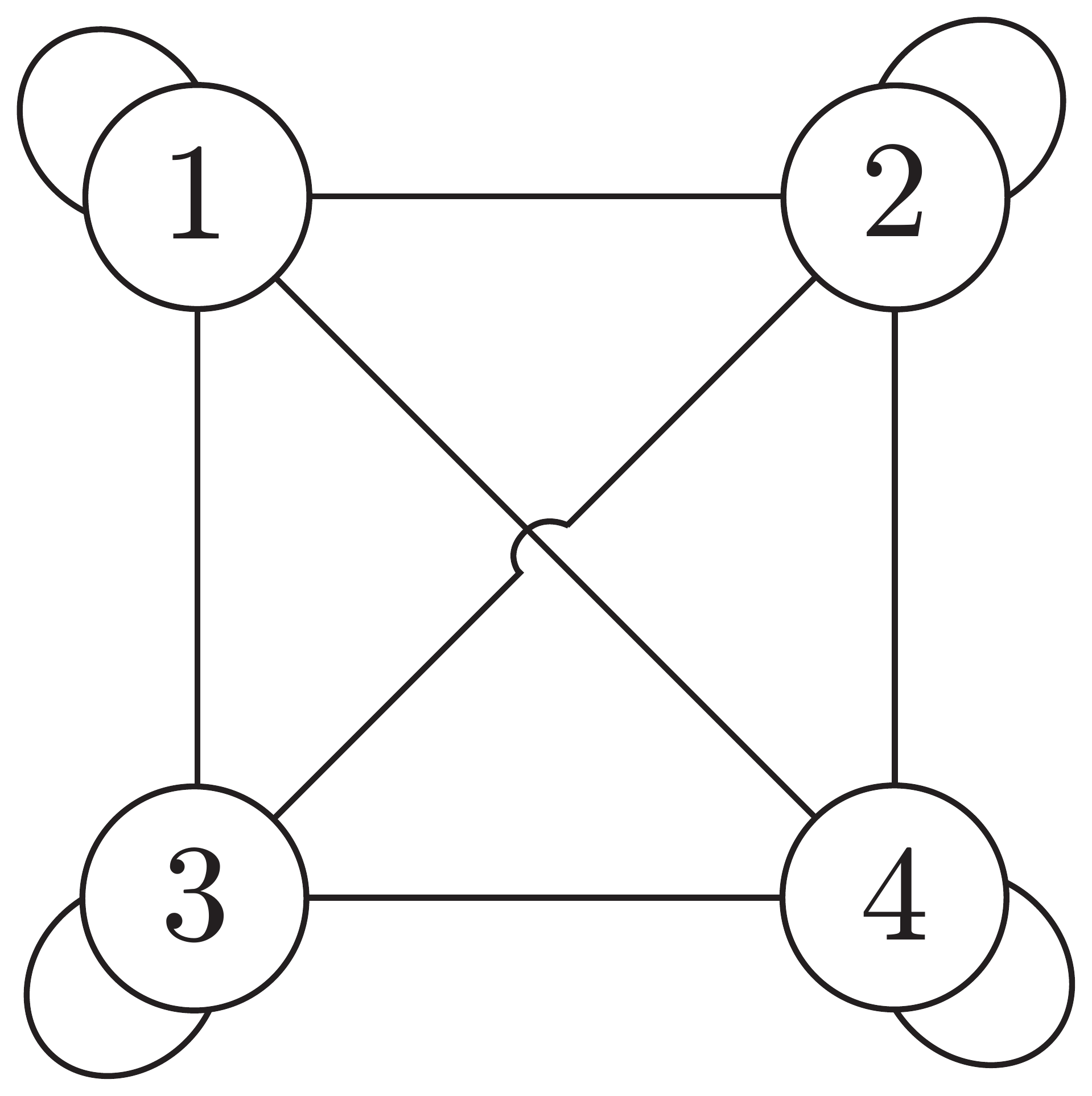}{The complete graph with 4 nodes}

From Eq.~\eqref{eq:trace} and $\vbf{M} \approx \vbf{C}$, $m$-th moment
$\ex{f^{(n)}_{\Lambda}}{(1-\Lambda)^m}$ is approximated by
\begin{align}
  \ex{f^{(n)}_{\Lambda}}{(1-\Lambda)^m} \approx \frac{1}{n-1} \ex{\prob{\vbf{\Gamma}}}{\tr{\vbf{C}^{m}}}.
\end{align}
As $n \rightarrow \infty$, the error of the approximation approaches
to $0$.
Hence, we investigate $\ex{\prob{\vbf{\Gamma}}}{\tr{\vbf{C}^{m}}}$ in
order to prove that $\ex{f^{(n)}_{\Lambda}}{(1-\Lambda)^m}$ converges
to Eqs.~\eqref{eq:wigner_even_moment} and \eqref{eq:wigner_odd_moment}
as $n \rightarrow \infty$.

To show the convergence of even moment $\ex{f^{(n)}_{\Lambda}}{(1-\Lambda)^{2m}}$, we
investigate $\ex{\prob{\vbf{\Gamma}}}{\tr{\vbf{C}^{2m}}}$.
Using the independence of $C(i, j)$, even moment
$\ex{\prob{\vbf{\Gamma}}}{\tr{\vbf{C}^{2m}}}$ is bounded by
\begin{align}
  \ex{\prob{\vbf{\Gamma}}}{\tr{\vbf{C}^{2m}}} &= \sum_{\vbf{v}_m \in \Phi_{n, 2m}} \prod_{l = 1}^{h(\vbf{v}_m)} \ex{\prob{\vbf{\Gamma}}}{C(e_l)^{m_l}} \nonumber\\
  \le \sum_{l = 0}^m |Z_{l, m}| & \frac{\ex{p_W}{W^{2}}^l\,\ex{p_W}{W^{2(m-l)}} \rho^l}{\ex{p_W}{W}^{2m}\omega_{\rm min}^{2m-2l}}  \left(1+o(1)\right) \nonumber\\
  \le \sum_{l = 0}^m |Z_{l, m}| & \frac{\ex{p_W}{W^{2}}^l \rho^l}{\ex{p_W}{W}^{2m}\omega_{\rm min}^{2m-2l}}  \left(1+o(1)\right)
  \label{eq:upper_trC2m},
\end{align}
where $Z_{l, m}$ is a set of cycles $(v_1, v_2, ... , v_{2m})$ with length
$2m$ and disjoint $l+1$ nodes.  Note that $Z_{l, m} \subset \Phi_{n, 2m}$.
For example, $(1, 2, 3, 1, 2, 3)$ is included in $Z_{2, 3}$.  The
cycles included in $Z_{l, m}$ are composed of at least $l$ disjoint
links.  To derive the second right-hand side of the above equation, we
first divide all cycles in $\Phi_{n, 2m}$ into the sets $Z_{l, m}$ of
the cycles that the number of disjoint links is $l$.  Each set has
$|Z_{l, m}|$ cycles.  Using Eq.~\eqref{eq:moment_cij}, for $\vbf{v}_m
\in Z_{l, m}$, we obtain
\begin{align}
  \prod_{l = 1}^{h(\vbf{v}_m)} \ex{\prob{\vbf{\Gamma}}}{C(e_l)^{m_l}} \le \frac{\ex{p_W}{W^{2}}^l\,\ex{p_W}{W^{2(m-l)}} \rho^l}{\ex{p_W}{W}^{2m}\omega_{\rm min}^{2m-2l}},
\end{align}
since that $\ex{p_W}{W^2}$ is the largest in $m$-th moments
$\ex{p_W}{W^m}$ for $m \ge 2$.
In the first right-hand side of Eq.~\eqref{eq:upper_trC2m}, the reason
why we do not consider the cycles longer than $m$ is as follows.
First, $\ex{\prob{\vbf{\Gamma}}}{C(i, j)} = 0$.  Hence, $\prod_{l =
  1}^{h(\vbf{v}_m)} \ex{\prob{\vbf{\Gamma}}}{C(e_l)^{m_l}}$ is 0
except if $m_l \ge 2$ for $1 \le l \le h(\vbf{v}_m)$.  Since $m_l \ge
2$ for $1 \le l \le h(\vbf{v}_m)$, the number of links $l$ in the
cycle is at most $m$.

To extract the main term of
$\ex{\prob{\vbf{\Gamma}}}{\tr{\vbf{C}^{2m}}}$, we rewrote Eq.~\eqref{eq:upper_trC2m} with
\begin{align}
  \ex{\prob{\vbf{\Gamma}}}{\tr{\vbf{C}^{2m}}} &\le \sum_{l = 1}^{m} \eta_{l,m} = \eta_{m,m} \sum_{l = 1}^{m} \frac{\eta_{l,m}}{\eta_{m,m}},
  \label{eq:upper_trC2m2}
\end{align}
where $\eta_{l,m}$ is 
\begin{align}
  \eta_{l,m}  &=  |Z_{l, m}| \frac{\ex{p_W}{W^{2}}^l \rho^l}{\ex{p_W}{W}^{2m}\omega_{\rm min}^{2m-2l}}   \label{eq:eta}.
\end{align}
In the above equation, $|Z_{l, m}|$ is given by
\begin{align}
  |Z_{l, m}| &= |Z_a(l, m)| \, |Z_b(l, m)| \, |Z_c(l, m)| \nonumber\\
                &= \frac{n !}{(n - l - 1)!} \comb{2m}{2l} (l+1)^{4\,(m - l)} \frac{1}{l+1}\comb{2l}{l},
  \label{eq:num_Z}  
\end{align}
where $Z_a(l, m)$ is the number of permutations $(i_k)_{1 \le k \le l+1}$ by
selecting $l+1$ nodes from $n$ nodes, and is given by
\begin{align}
  |Z_a(l, m)| = \frac{n !}{(n - l - 1)!}.
\end{align}
Then, $|Z_b(l, m)|$ is the number of combinations that each $i_k$
appears more than once in the cycle with length $2m$, and is given by
\begin{align}
  |Z_b(l, m)| \!&=\! \comb{2m}{2l} \!\!\left((l+1)^2\right)^{2(m - l)} \!\!=\! \comb{2m}{2l} \!\!\left(l+1\right)^{4(m - l)}.
\end{align}
In the above equation, since there is no restriction except that each
$i_k$ must appear at least twice, it is multiplied by
$\left(l+1\right)^{4(m - l)}$.
Moreover, $Z_c(l, m)$ is the number of the second appearance positions for
$i_k$, and is given by
\begin{align}
  |Z_{c}(l, m)| &= \frac{1}{l+1}\comb{2l}{l}.
\end{align}
Note that the right side of the above equation is the Catalan number.

By substituting Eq.~\eqref{eq:num_Z} into Eq.~\eqref{eq:eta},
$\eta_{l, m}$ is given by
\begin{align}
  \eta_{l,m}  &=  |Z_{l, m}| \frac{\ex{p_W}{W^{2}}^l \rho^l}{\ex{p_W}{W}^{2m}\omega_{\rm min}^{2m-2l}}  \nonumber\\
  &= \frac{n !}{(n - l - 1)!} \comb{2m}{2l} (l + 1)^{4(m - l)}  \nonumber\\
  & \hspace{2cm} \frac{1}{l + 1} \comb{2l}{l}\frac{\ex{p_W}{W^{2}}^l \rho^l}{\ex{p_W}{W}^{2m} \omega_{\rm min}^{2m-2l}}.
\end{align}
Then, $\frac{\eta_{l,m}}{\eta_{m,m}}$ in~\eqref{eq:upper_trC2m2} is bounded by
\begin{align}
  \frac{\eta_{l,m}}{\eta_{m,m}} &= \frac{(n - m - 1)!}{(n - l - 1!)} \comb{2m}{2l} (l+1)^{4(m-l)}  \nonumber\\
  & \hspace{1.0cm} \frac{m+1}{l+1} \frac{\comb{2l}{l}}{\comb{2m}{m}} \frac{\ex{p_W}{W^{2}}^l\ex{p_W}{W^{2m-2l}} \rho^l}{\omega_{\rm min}^{2m-2l} \ex{p_W}{W^2}^m \rho^m} \nonumber\\
  &\le \comb{2m}{2l} \frac{n^{l-m} (l+1)^{4(m-l)} 4^{l-m}}{\omega_{\rm min}^{2(m-l)} \ex{p_W}{W^2}^{m-l} \rho^{m-l}} \nonumber\\
  &\le \frac{n^{l-m}\,2\,m^{2(m-l)}\,m^{4(m-l)} 4^{l-m}}{\omega_{\rm min}^{2(m-l)} \ex{p_W}{W^2}^{m-l} \rho^{m-l}} \nonumber\\
  &\le 2\left[\frac{\omega_{\rm avg}\,m^6}{4\,\omega_{\rm min}^2\ex{p_W}{W^2}}\right]^{m-l}.
  \label{eq:upper_eta_ratio}
\end{align}
To obtain the upper bound of $\frac{\eta_{l,m}}{\eta_{m,m}}$, we used
the Stirling's approximation, which is
\begin{align}
  &\sqrt{2\,\pi} n^{n+1/2}\,e^{-n} \le n! \le n^{n+1/2}\,e^{-n+1}.
\end{align}
According to the following discussion, if the degree condition is
satisfied the right-hand side of Eq.~\eqref{eq:upper_eta_ratio} for $m
> l$ becomes $o(1)$.  First, for $m > l$, we derive
\begin{align}
  \left[\frac{\omega_{\rm avg}\,m^6}{4\,\omega_{\rm min}^2\ex{p_W}{W^2}}\right]^{m-l}  = o(1).
  \label{eq:right_order}
\end{align}
To prove the above equation, we should show
\begin{align}
  \frac{\omega_{\rm avg}\,m^6}{4\,\omega_{\rm min}^2\ex{p_W}{W^2}} = o(1).
  \label{eq:cond_m}
\end{align}
Similarly to~\cite{Chung03:Spectra}, using $m = \log n$, we obtain condition
\begin{align}
  \omega_{\rm min}^2 = \Omega((\log n)^6) \frac{\omega_{\rm avg}}{\ex{p_W}{W^2}},
  \label{eq:cond_min}
\end{align}
to satisfy Eqs.~\eqref{eq:right_order} and \eqref{eq:cond_m}.
The condition~\eqref{eq:cond_min} is the same as the degree
condition~\eqref{eq:cond_degree}.

Therefore, if the degree condition~\eqref{eq:cond_degree} is
satisfied, for $m > l$, we obtain
\begin{align}
  \frac{\eta_{l,m}}{\eta_{m,m}}  = o(1).
\end{align}
By substituting the above equation into Eq.~\eqref{eq:upper_trC2m2},
the upper bound of $\ex{\prob{\vbf{\Gamma}}}{\tr{\vbf{C}^{2m}}}$ is given by
\begin{align}
  \ex{\prob{\vbf{\Gamma}}}{\tr{\vbf{C}^{2m}}} \le \left(1 + o(1)\right) \eta_{m,m}.
\end{align}

On the other hand, the lower bound of
$\ex{\prob{\vbf{\Gamma}}}{\tr{\vbf{C}^{2m}}}$ is given by
\begin{align}
  &\ex{\prob{\vbf{\Gamma}}}{\tr{\vbf{C}^{2m}}} = \sum_{\vbf{v}_m \in \Phi_{n, m}} \prod_{h = 1}^{h(\vbf{v}_m)} \ex{\prob{\vbf{\Gamma}}}{C(e_l)^{m_l}} \nonumber\\
  &\hspace{1cm} \ge |Z_{m,m}| \ex{\prob{\vbf{\Gamma}}}{C(i, j)^2}^m  \nonumber\\
  &\hspace{1cm} \ge |Z_{m,m}| \rho^m \frac{\ex{p_W}{W^2}^m}{\ex{p_W}{W}^{2m}} (1 - o(1)) \nonumber\\
  &\hspace{1cm} =  (1 - o(1))\,\eta_{m,m}.
  \label{eq:lower_trC2m}
\end{align}
Note that we obtained the second right-hand side of the above equation
by only using the cycles with disjoint $m$ links.
To derive the third right-hand side, we used
\begin{align}
  \ex{\prob{\vbf{\Gamma}}}{C(i, j)^2} \ge \rho \frac{\ex{p_W}{W^2}}{\ex{p_W}{W}^2} (1 - o(1)).
\end{align}
This can be derived from Eq.~\eqref{eq:moment_cij_pre}.

According to Eqs.~\eqref{eq:upper_trC2m2} and \eqref{eq:lower_trC2m},
$\ex{\prob{\vbf{\Gamma}}}{\tr{\vbf{C}^{2m}}}$ is bounded by
\begin{align}
  \eta_{m,m} (1 - o(1)) \le \ex{\prob{\vbf{\Gamma}}}{\tr{\vbf{C}^{2m}}} \le \eta_{m,m} (1 + o(1)).
\end{align}
As $n \rightarrow \infty$, even moment $\ex{f_{\Lambda}^{(n)}}{(1-\lambda)^{2m}}$ converges to
\begin{align}
  &\lim_{n \rightarrow \infty} \ex{f_{\Lambda}^{(n)}}{(1-\lambda)^{2m}} = \lim_{n \rightarrow \infty} \frac{1}{n-1} \ex{\prob{\vbf{\Gamma}}}{\tr{\vbf{C}^{2m}}} \nonumber\\
  &\hspace{0.1cm} = \lim_{n \rightarrow \infty} \frac{1 \pm o(1)}{n-1} \eta_{m,m} \nonumber\\
  &\hspace{0.1cm} = \lim_{n \rightarrow \infty} \frac{n!}{\!(n-1)(n-m-1)!}\frac{1}{m+1} \!\comb{2m}{m}\! \frac{\ex{p_W}{W^2}^m\!\!\rho^m}{\ex{p_W}{W}^{2m}} \nonumber\\
  &\hspace{0.1cm} = \lim_{n \rightarrow \infty} g(n) \frac{\ex{p_W}{W^2}^m}{\omega_{\rm avg}^m\ex{p_W}{W}^{2m}} \frac{1}{m+1}\comb{2m}{m} \nonumber\\
  &\hspace{0.1cm} = \frac{\ex{p_W}{W^2}^m}{\omega_{\rm avg}^m \ex{p_W}{W}^{2m}} \frac{1}{m+1}\comb{2m}{m} \nonumber\\
  &\hspace{0.1cm} = \left(\frac{\tilde{r}}{2}\right)^{2m} \frac{(2\,m) !}{m !\,(m+1)!},
\end{align}
where $g(n)$ is defined by
\begin{align}
  g(n) :=  \frac{n!}{(n\!-\!1)(n \!-\!m\!-\!1)! \,n^m}.
\end{align}
Note that $\lim_{n \rightarrow \infty} g(n) = 1$.

Therefore, if weighted network $G$ satisfies the degree
condition~\eqref{eq:cond_degree}, even moment
$\ex{\prob{\vbf{\Gamma}}}{\tr{\vbf{C}^{2m}}}$ converges to
$\ex{\tilde{f}_{\Lambda}}{(1-\lambda)^{2m}}$ as $n \rightarrow
\infty$.

Next, we investigate odd moment
$\ex{\prob{\vbf{\Gamma}}}{\tr{\vbf{C}^{2m+1}}}$.  Similarly to even
moment $\ex{\prob{\vbf{\Gamma}}}{\tr{\vbf{C}^{2m}}}$, $\ex{\prob{\vbf{\Gamma}}}{\tr{\vbf{C}^{2m+1}}}$ is bounded by
\begin{align}
  \ex{\prob{\vbf{\Gamma}}}{\tr{\vbf{C}^{2m+1}}} &= \sum_{\vbf{v}_m \in \Phi_{n, 2m+1}} \prod_{l = 1}^{h(\vbf{v}_m)} \ex{\prob{\vbf{\Gamma}}}{C(e_l)^{m_l}} \nonumber\\
  \le \sum_{l = 0}^m |Z_{l, m}| & \frac{\ex{p_W}{W^{2}}^{l-1} \rho^l}{\ex{p_W}{W}^{2m+1}\omega_{\rm min}^{2m-2l+1}} \left(1+o(1)\right) \nonumber\\
  \le \sum_{l = 0}^m \frac{\eta_{l, m}'}{\omega_{\rm min}}  & =  \frac{\eta_{m, m}'}{\omega_{\rm min}} \sum_{l = 0}^m \left(\frac{\eta_{l, m}'}{\eta_{m, m}'} \right),
\end{align}
where $\sum_{h=1}^l m_h = 2m+1$, and $\eta_{l, m}'$ is given by
\begin{align}
  \eta_{l, m}' = |Z_{l, m}| & \frac{\ex{p_W}{W^{2}}^{l-1}  \rho^l}{\ex{p_W}{W}^{2m+1}\omega_{\rm min}^{2m-2l}} \left(1+o(1)\right).
\end{align}
Using the similar way of the even moment, $\eta_{l, m}'/\eta_{m, m}'$ is bounded by
\begin{align}
  \frac{\eta_{l, m}'}{\eta_{m,m}'} &\le 2\left[\frac{\omega_{\rm avg}\,m^6}{4\,\omega_{\rm min}^2\ex{p_W}{W^2}}\right]^{m-l}.
  \label{eq:upper_etap_ratio}
\end{align}
If the degree condition~\eqref{eq:cond_degree} is satisfied, for $m > l$, we obtain 
\begin{align}
  \left[\frac{\omega_{\rm avg}\,m^6}{4\,\omega_{\rm min}^2}\right]^{m-l}  = o(1).
\end{align}
Hence, the upper bound of $1/(n-1) \ex{\prob{\vbf{\Gamma}}}{\tr{\vbf{C}^{2m+1}}}$ is given by
\begin{align}
  \frac{1}{n-1} \ex{\prob{\vbf{\Gamma}}}{\tr{\vbf{C}^{2m+1}}} \le \frac{1 + o(1)}{n-1} \frac{\eta_{m,m}'}{\omega_{\rm min}}.
  \label{eq:upper_trC2m1}
\end{align}
As $n \rightarrow \infty$, the right-hand side of the above equation converges to 
\begin{align}
  &\lim_{n \rightarrow \infty} \frac{1 + o(1)}{n-1} \frac{\eta_{m,m}'}{\omega_{\rm min}} \nonumber\\
  &\hspace{0.1cm} = \lim_{n \rightarrow \infty}\frac{g(n)}{m+1} \!\comb{2m}{m}\! \frac{\ex{p_W}{W^2}^{m-1}}{\ex{p_W}{W}^{2m+1}\omega_{\rm avg}^m\omega_{\rm min}} \nonumber\\
  &\hspace{0.1cm} = \left(\frac{\tilde{r}}{2}\right)^{2m}\!\!\!\frac{(2\,m) !}{m !\,(m+1)!} \lim_{n \rightarrow \infty}\frac{g(n)}{\ex{p_W}{W^2}\ex{p_W}{W}\omega_{\rm min}} \nonumber\\
  &\hspace{0.1cm} = 0.
\end{align}
Note that $1/\omega_{\rm min} = o(1)$ since $\omega_{\rm min}$ is a
increasing function of $n$.  Hence, as $n\rightarrow \infty$, odd
moment $\ex{f^{(n)}_{\Lambda}}{(1-\Lambda)^{2m+1}}$ is given by
\begin{align}
  \lim_{n \rightarrow \infty} \ex{f^{(n)}_{\Lambda}}{(1-\Lambda)^{2m+1}} &= \lim_{n \rightarrow \infty} \frac{(1 + o(1))}{n-1} \frac{\eta_{m,m}'}{\omega_{\rm min}} \nonumber\\
  & =0.
\end{align}

Therefore, if weighted network $G$ satisfies the degree
condition~\eqref{eq:cond_degree}, odd moment also converges to
$\ex{\tilde{f}_{\Lambda}}{(1-\Lambda)^{2m+1}}$ as as $n \rightarrow
\infty$.
\end{proof}

\section{Numerical Example}
\label{sec:example}

We confirm the the validity of the degree
condition~\eqref{eq:cond_degree} and Eq.~\eqref{eq:radius} in the
Wigner's semicircle law for weighted network $G$ derived in
Sect.~\ref{sec:analysis}.  For the sake of space, we only show the
results using the BA model~\cite{Barabasi99:model}, which is a famous
random network generation model.

Similarly to~\cite{Sakumoto19:IEICE}, we generate weighted network $G$ with the
following procedures based on the BA model.
\begin{enumerate}
\item Generate an unweighted network with $n$ nodes and average degree
  $\omega_{\rm avg}^{\rm BA}$ according to the BA model.
\item Randomly cut the links of the unweighted network until the
  average degree is $\omega_{\rm avg}$.  This procedure prevents
  minimum degree $\omega_{\rm min}$ from being fixed, and does not
  lose the scale-free property of the BA network~\cite{Sakumoto19:IEICE}.
\item Randomly assign the weight of each link using the Pareto
  distribution $p_W^{\rm par}(w)$ with $\ex{p_W}{W} = 1$.  Note that
  $p_W^{\rm par}(w) \propto \alpha\,w^{-\alpha-1}$.
\item Divide the weight of each link by maximum link weight $w_{\rm
  max}$.  By performing this procedure, the link weight is smaller
  than or equal to $1$, and its distribution satisfies the condition
  of $p_W(w)$ used in Section 3.  Note that this procedure does not
  change normalized Laplacian matrix $\vbf{N}$.
\end{enumerate}

In order to confirm the validity of the degree
condition~\eqref{eq:cond_degree} in the Wigner's semicircle law for
weighted network $G$, we compare eigenvalue density
$f_\Lambda^{(n)}(\lambda)$ of the normalized Laplacian matrix
$\vbf{N}$ and semicircle distribution $\tilde{f}_{\Lambda}(\lambda)$.
To evaluate the difference of them, we use the relative error
$\epsilon_{\rm d}$, which is defined by
\begin{align}
  \epsilon_{\rm d} := \frac{1}{n_h} \sum_{i=1}^{n_h} \frac{|F_n(\theta_i) - F^*(\theta_i)|}{F(\theta_i)},
  \label{eq:dist_err}
\end{align}
where $\theta_i = (i-1/2)h_b + \lambda_2$ and $h_b = (\lambda_n -
\lambda_2)/n_h$.  In~\eqref{eq:dist_err}, $F_n(\theta_i)$ is the value
obtained with dividing the number of eigenvalues of $\vbf{N}$ within
$[\theta_i - h_b/2, \theta_i + h_b/2]$ by $n-1$.  On the contrary,
$F^*(\theta_i)$ is the value of the integral of
$\tilde{f}_{\Lambda}(\lambda)$ within $[\theta_i - h_b/2, \theta_i +
  h_b/2]$.

In order to confirm the validity of Eq.~\eqref{eq:radius}, we compare
spectral radius $r$ of normalized Laplacian matrix $\vbf{N}$ and
$\tilde{r}$ calculated by Eq.~\eqref{eq:radius}.  For these
comparisons, we use relative error $\epsilon_r$, which is defined by
\begin{align}
  \epsilon_r := \frac{|\tilde{r} - r|}{r}.
  \label{eq:r_err}
\end{align}

In the numerical example, we use the parameter configuration shown in
Tab.~\ref{tab:param} as a default parameter configuration.

\begin{table}[t]
  \caption{Parameter configuration}
  \begin{center}
    \begin{tabular}{l|c} 
      \hline
      Average degree of unweighted BA networks, $\omega_{\rm avg}^{(BA)}$ & 40\\
      Number of nodes of weighted network $G$, $n$ & 1,000\\
      Average degree of weighted network $G$, $\omega_{\rm avg}$ & 20\\
      Pareto index of the Pareto distribution, $\alpha$ & 3 \\
      Number of bins, $n_h$ & 50 \\
      \hline
    \end{tabular}
    \label{tab:param}
  \end{center}
\end{table}

First, we confirm the relationship between the characteristics of
weighted network $G$ generated in the above procedure and the degree
condition~\eqref{eq:cond_degree}.
Figure~\ref{fig:avg-deg_vs-min-deg_ba_vs_er_N=1000_k=40} shows the
square of the minimum degree, $\omega_{\rm min}^2$ for different
average degree $\omega_{\rm avg}$.  In this figure, we also plots the
results with the average order $\omega_{\rm avg}$ on the $y$ axis for
comparison.  From these results, as $\omega_{\rm avg}$ increases, the
difference between $\omega_{\rm min}^2$ and $\omega_{\rm avg}$
increases, and hence the degree condition~\eqref{eq:cond_degree} is
more easily satisfied.  Figure~\ref{fig:alpha_vs_w2} shows second
moment $\ex{p_W}{W^2}$ of link weights for different Pareto index
$\alpha$.  From this figure, as $\alpha$ increases, $\ex{p_W}{W^2}$
increases, and hence the degree condition~\eqref{eq:cond_degree} is
also more easily satisfied.

\insertfig{avg-deg_vs-min-deg_ba_vs_er_N=1000_k=40}{Average degree  $\omega_{\rm avg}$ vs. the square of the minimum degree, $\omega_{\rm min}^2$}
\insertfig{alpha_vs_w2}{Second moment $\ex{p_W}{W^2}$ for different Pareto index $\alpha$}

Figures~\ref{fig:dist-eig_ba}~(a) and (b) show eigenvalue density
$f_\Lambda^{(n)}(\lambda)$ of the normalized Laplacian matrix
$\vbf{N}$~(i.e, $F_n(\theta)$) and semicircle distribution
$\tilde{f}_{\Lambda}(\lambda)$.  For reference, we show the results of
all link weights $w(i, j) = 1$ in Fig.~\ref{fig:dist-eig_ba}~(c).
According to the results, eigenvalue density
$f_\Lambda^{(n)}(\lambda)$ for $\alpha = 5$ is closer to semicircle
distribution $\tilde{f}_{\Lambda}(\lambda)$ than that for $\alpha =
3$.  This is consistent with the result of $\ex{p_W}{W^2}$ shown in
Fig.~\ref{fig:alpha_vs_w2}.  Hence, we visually confirm the validity
of the degree condition~\eqref{eq:cond_degree}.

\insertfigsubthree{dist-eig_ba_a=3}{pareto distribution with $\alpha =
  3$} {dist-eig_ba_a=5}{pareto distribution with $\alpha =
  5$}{dist-eig_ba}{$w(i, j) = 1$} {dist-eig_ba}{Eigenvalue density
  $f_\Lambda^{(n)}(\lambda)$ of the normalized Laplacian matrix
  $\vbf{N}$~(i.e, $F_n(\theta)$) and semicircle distribution
  $\tilde{f}_{\Lambda}(\lambda)$}

Figure~\ref{fig:err-dist-eig} shows relative error $\epsilon_{\rm d}$
of eigenvalue density $f_{\Lambda}^{(n)}(\lambda)$ for different
average degree $\omega_{\rm avg}$.  In this figure, we also show the
result for all link weights $w(i, j) = 1$ for reference.  Since $n$ is
finite, relative error $\epsilon_{\rm d}$ is not $0$.  We assume that
relative error $\epsilon_{\rm d}$ is almost equal to the result for
$w(i, j) = 1$, $f_{\Lambda}^{(n)}(\lambda)$ converges to
$\tilde{f}_{\Lambda}(\lambda)$ as $n\rightarrow \infty$.  From the
results in Fig.~\ref{fig:err-dist-eig}, relative error $\epsilon_{\rm
  d}$ decreases as average degree $\omega_{\rm avg}$ increases or
Pareto index $\alpha$ increases.  The result is consistent with the
results shown in
Figs.~\ref{fig:avg-deg_vs-min-deg_ba_vs_er_N=1000_k=40} and
\ref{fig:alpha_vs_w2}.  Hence, the degree
condition~\eqref{eq:cond_degree} is valid.  Moreover, relative error
$\epsilon_{\rm d}$ for $\alpha = 5$ is almost same as the that for
$w(i, j) = 1$.  Hence, if $\alpha \ge 5$, $f_{\Lambda}^{(n)}(\lambda)$
follows the Wigner's semicircle law.

\insertfig{err-dist-eig}{Relative error $\epsilon_{\rm d}$ of
  eigenvalue density $f_{\Lambda}^{(n)}(\lambda)$ for different
  average degree $\omega_{\rm avg}$}

Figure~\ref{fig:r-eig} shows spectral radius $r$ and $\tilde{r}$
calculated from Eq.~\eqref{eq:radius} for different average degree
$\omega_{\rm avg}$.  In this figure, we also plot the result for all
link weights $w(i, j) = 1$ for reference.  From Fig.~\ref{fig:r-eig},
spectral radius $r$ almost coincides with $\tilde{r}$ except for
$\alpha = 3$. Hence, spectral radius $r$ can be calculated accurately
using Eq.~\eqref{eq:radius} if the degree
condition~\eqref{eq:cond_degree} is satisfied.

\insertfig{r-eig}{Spectral radius $r$ and $\tilde{r}$ calculated from
  Eq.~\eqref{eq:radius} for different average degree $\omega_{\rm
    avg}$}

Figure~\ref{fig:r_err-eig} shows relative error $\epsilon_r$ of
$\tilde{r}$ calculated by Eq.~\eqref{eq:radius} for different
average degree $\omega_{\rm avg}$.  In this figure, the result of for
all link weights $w(i, j) = 1$ is also plotted for reference.
According to Fig.~\ref{fig:r_err-eig}, it is clear that relative error
$\epsilon_r$ is small when the degree condition~\eqref{eq:cond_degree}
is easily satisfied as with the result shown in
Fig.~\ref{fig:err-dist-eig}.  Hence, Eq.~\eqref{eq:radius} is
valid as the approximate expression of spectral radius $r$.

\insertfig{r_err-eig}{Relative error $\epsilon_r$ of
$\tilde{r}$ calculated by Eq.~\eqref{eq:radius} for different
average degree $\omega_{\rm avg}$}

From the above results, we conclude that the degree
condition~\eqref{eq:cond_degree} and Eq.~\eqref{eq:radius} derived in
Sect.~\ref{sec:analysis} are valid.

\section{Conclusion and Future Work}
\label{sec:conclusion}

In this paper, we have clarified the Wigner's semicircle law for
weighted network $G$ on the basis of the random matrix theory.  This
law indicates that if $G$ with $n$ nodes satisfies the degree
condition~\eqref{eq:cond_degree}, the eigenvalue density of the
normalized Laplacian matrix $\vbf{N}$ converges to the semicircle
distribution determined by the approximated spectral radius
$\tilde{r}$ as $n \rightarrow \infty$.  By using
Eq.~\eqref{eq:radius}, we can calculate $\tilde{r}$ from a few network
statistics~(the average degree, the average link weight, and the
square average link weight).  Hence, the eigenvalue distribution of
$\vbf{N}$ can be obtained from these network statistics without giving
all matrix elements $N(i, j)$ accurately.  Our results provide a new
analysis method for weighted network $G$ using the spectral graph
theory and the random graph theory.

As future work, we are planing to analyze and design actual networks
using the Wigner's semicircle law clarified in this paper.  In
particular, we will investigate the characteristics of the information
dissemination on the social network, and design the social media to
control the speed of the information dissemination.

\section*{Acknowledgement}
  
This work was supported by JSPS KAKENHI Grant Number 17H01737.

\bibliographystyle{IEEEtran}
\bibliography{bib/complex_model,bib/random_matrix,bib/misc,bib/sakumoto,bib/spectral,bib/network_analysis}

\profile{Yusuke Sakumoto} {received M.E. and Ph.D. degrees in the
  Information and Computer Sciences from Osaka University in 2008 and
  2010, respectively.  From 2010 to 2019, he was a associate professor
  of Tokyo Metropolitan University.  He is currently an associate
  professor at Kwansei Gakuin University.  His research work is in the
  area of communication network, electricity network, and social
  network.  He is a member of the IEEE, IEICE and IPSJ.}

\profile{Masaki Aida} {received his B.S. degree in Physics and
  M.S. degree in Atomic Physics from St.~Paul's University, Tokyo,
  Japan, in 1987 and 1989, respectively, and his Ph.D. in
  Telecommunications Engineering from the University of Tokyo, Japan,
  in 1999.  In April 1989, he joined NTT Laboratories.  From April
  2005 to March 2007, he was an Associate Professor at Tokyo
  Metropolitan University.  He has been a Professor of the Graduate
  School of Systems Design, Tokyo Metropolitan University since April
  2007.  His current interests include analysis of social network
  dynamics and distributed control of computer communication networks.
  He received the Best Tutorial Paper Award and the Best Paper Award
  of IEICE Communications Society in 2013 and 2016, respectively, and
  IEICE 100-Year Memorial Paper Award in 2017.  He is a fellow of
  IEICE and a member of the IEEE, ACM and ORSJ.}

\end{document}